\documentclass[conference]{IEEEtran}

\usepackage{graphics}
\usepackage{epstopdf}
\usepackage[pdftex]{graphicx}
\usepackage{stfloats}
\usepackage{array}
\usepackage{float}
\usepackage[cmex10]{amsmath}
\usepackage{amsmath,amsfonts,amssymb}
\usepackage{graphicx,graphics}
\usepackage{enumerate}
\usepackage{color}
\usepackage{verbatim}
\usepackage{amsthm}
\usepackage{multirow}
\usepackage{subcaption}
\usepackage{siunitx}
\usepackage{amsmath,amsthm}
\usepackage{cleveref}
\usepackage{cite}
\usepackage[font=footnotesize]{caption}
\usepackage{fixltx2e}
\usepackage[USenglish]{babel}

\usepackage{geometry}

\newcommand{\InfFunc}{{{\cal{I}}^{\boldmath \beta_+(t)}_{\boldmath \beta_-(t)}}}

\begin{document}

%\copyrightstatement

\title{Impulsive Noise Mitigation in Underwater Acoustic Communication Systems: Experimental Studies }

\author{
Reza Barazideh$^{\dag}$, Wensheng Sun$^{\ddag}$, Balasubramaniam Natarajan$^{\dag}$, Alexei V. Nikitin$^{\dag \ast}$, and Zhaohui Wang$^{\ddag}$ \\
\small $^{\dag}$ Department of Electrical and Computer Engineering, Kansas State University, Manhattan, KS, USA.\\
$^{\ddag}$ Department of Electrical and Computer Engineering, Michigan Technological University, Houghton, MI, USA.\\
$^{\ast}$ Nonlinear LLC, Wamego, KS, USA.\\
Emails:\{rezabarazideh, bala\}@ksu.edu, \{wsun3, zhaohuiw\}@mtu.edu, avn@nonlinearcorp.com
\thanks{978-1-7281-0554-3/19/\$31.00 \copyright 2019 IEEE}

}

%\author{
%\begin{tabular}{c} Roohollah~Amiri \\ Electrical and Computer Engineering \\ Boise State University, Idaho, USA\\ roohollahamiri@u.boisestate.edu \end{tabular} \and
%\begin{tabular}{c} Hani~Mehrpouyan \\ Electrical and Computer Engineering \\ Boise State University, Idaho, USA\\ hanimehrpouyan@boisestate.edu \end{tabular}
%\thanks{978-1-7281-0554-3/19/\$31.00©2019 IEEE}}

\maketitle

\begin{abstract}

Impulsive noise is a major impediment to orthogonal frequency-division multiplexing (OFDM) based underwater acoustic (UWA) communications. In this work, we evaluate the performance of a memoryless analog nonlinear preprocessor (MANP) that is used to mitigate outliers. The proposed MANP exhibits intermittent nonlinearity only in the presence of the impulsive noise and suppresses the power of outliers based on their amplitudes. Since the outliers are distinguishable in the analog domain prior to anti-aliasing filtering, the MANP outperforms its digital counterparts in all scenarios. Experimental results using data collected in an under-ice environment, demonstrate the superior BER performance of our approach relative to classical nonlinear approaches such as blanking and clipping.

 %before their detrimental effects come to the linear digital processing chain of the conventional receiver.

\end{abstract}

\begin{IEEEkeywords}
Impulsive noise, memoryless analog nonlinear preprocessor (MANP), orthogonal frequency-division multiplexing (OFDM), underwater acoustic (UWA) communications.
\end{IEEEkeywords}

\IEEEpeerreviewmaketitle
\section{Introduction}

%Due to the low attenuation of sound,
Underwater acoustic (UWA) communication has been the most widely used technique for transmission in shallow water environments~\cite{RecentAdvancesinUWA_Chitre_2008,UWAModel_Chitre_2007}. The UWA communications are subject to multipath propagation with long delay spreading and strong Doppler effect \cite{Stojanovic_UWA_Channel_2009, Sun_Online_modelingUWA_2018}. In addition, impulsive noise is the main channel impairment in some underwater environments. For example, snapping shrimp noise in shallow warm waters~\cite{ShrimpNoise_Chitre_2005}, manmade noise near the shores~\cite{ManmadeUWAnoise_Hildebrand_2009}, and ice-cracking noise in Arctic environments~\cite{Ice_Cracking_Dwyer} are the common examples of impulsive noise. With the increasing demand for high data rate applications such as environmental monitoring, sonar, and communication between underwater vehicles, modern UWA communication systems have higher bandwidth. Since impulsive noise is typically wide band, it affects certain broadband modulation techniques such as orthogonal frequency-division multiplexing (OFDM) which is widely used in UWA communication. It is also widely known that impulsive noise is non-Gaussian and special care should be taken during the decoding and detection process~\cite{ShrimpNoise_Chitre_2005, OptNopt_Chitre_2006}. Thus, impulsive noise mitigation will positively impact the performance of UWA communication systems.

In prior literature, there are many approaches that have been proposed to mitigate impulsive noise. In general, impulsive noise mitigation techniques in OFDM systems can be divided into two classes. In the first class, the sparsity of the impulsive noise and the structure of the OFDM signal are exploited \cite{OptNopt_Chitre_2006}. In this class, the impulsive noise is first estimated based on the null and/or pilot subcarriers, and then the estimated impulsive noise is subtracted from the received signals. For example, compressive sensing (CS) techniques~\cite{CS_2008}, and  sparse Bayesian learning (SBL)~\cite{Lin13impulsive_SparseBayesian} fall in this class. In the second class, the high amplitude and the short duration of impulsive noise is exploited. The temporal structure of outliers guided the development various memoryless nonlinear approaches such as clipping and blanking which are the most common methods in this class~\cite{Zhidkovn08_Simpleanalysis}. Moreover, multiplethreshold blanking/clipping~\cite{MultiThershold_Rozic_2018}, and deep clipping~\cite{DeepCliping_2014_Juwono} are proposed to improve the performance of blanking and clipping at the cost of additional computational complexity. As shown in \cite{Zhidkovn08_Simpleanalysis}, the performance of all these methods degrades dramatically in severe impulsive environments.

Bandwidth reduction in the process of analog-to-digital conversion (ADC) is the main drawback of all these digital nonlinear approaches \cite{Khodam_Latincom, Khodam_ICC,Khodam_TVT}. To overcome this drawback, we proposed an Adaptive Nonlinear Differential Limiter (ANDL) to improve the bit-error-rate (BER) performance of uncoded OFDM-based communication systems in an additive noise channel \cite{Khodam_Latincom,Khodam_TVT}. A practical implementation of Adaptive Canonical Differential Limiter (ACDL) is studied in \cite{Khodam_ICC} to compensate for the impulsive noise in OFDM-based powerline communication (PLC) systems.

In this paper, for the first time, we investigate the performance of a memoryless analog nonlinear preprocessor (MANP) in a practical OFDM-based UWA communication system. The proposed MANP offers a compromise between clipping and blanking in response to the impulsivity level in the analog domain. The potency of the proposed MANP is evaluated based on the real data collected in Portage Lake, Michigan. We compare our proposed approach with conventional methods such as blanking and clipping and highlight the superiority of the MANP in the impulsive noise suppression. Experimental results show the improvement in BER performance, due to the fact that, unlike classical impulsive noise mitigation methods, MANP is implemented in the analog domain where the outliers are still distinct.

The remainder of this paper is organized as follows. Section~\ref{sec:System_Model} describes the UWA communication system model. Section~\ref{sec:Receiver Design} details the proposed receiver structure. Section~\ref{sec:Experimental Results} presents experimental results and Section~\ref{sec:Conclusion} draws the conclusions.

\section{System Model}\label{sec:System_Model}

\begin{figure*}
\centering
\includegraphics[scale=.38]{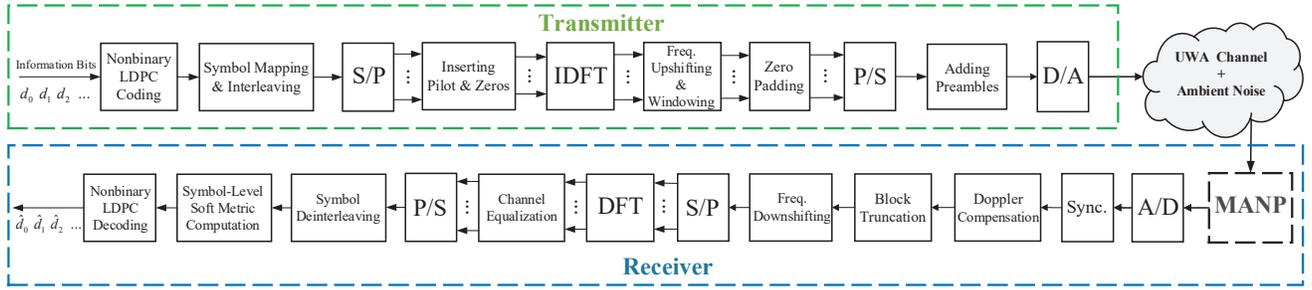}
\caption{System block diagram.}
\label{fig:System Model}
\vspace{-.15cm}
\end{figure*}

A simplified block diagram of the zero-padded OFDM-based UWA system is shown in Fig.~\ref{fig:System Model} and more details can be found in~\cite{Zhou_UWA_Book_2014}. At the transmitter, the information bits are encoded by nonbinary low-density parity-check (LDPC) codes. Symbols are mapped from the coded bits according to the desired modulation scheme and then interleaved. After inserting pilot symbols and zeros, the data are passed through an inverse discrete Fourier transform (IDFT) module to generate OFDM modulated baseband signals. Zero-padding is performed to counteract multipath effects after the signals are upshifted to the passband. Lastly, preambles are added to assist signal detection and synchronization.

%A simplified block diagram of the OFDM-based UWA system considered in this work is shown in Fig.~\ref{fig:System Model} and more details can be found in~\cite{Zhou_UWA_Book_2014}. At the transmitter, information bits are channel coded and then the encoded bits are interleaved. Subsequently, the interleaved data is modulated and passed through an inverse discrete Fourier transform (IDFT) module to generate OFDM symbols over orthogonal subcarriers. Zero padding (ZP) at the end of each OFDM symbol is used to counter multipath effects.

Let $T$ and $T_g$ denote the OFDM symbol duration and the length of the guard interval, respectively. The subcarrier spacing is $\Delta f{=}1/T$ and the total OFDM block duration is $T_{\rm{bl}}{=}T+T_g$. Therefore, an OFDM block with $N$ subcarriers has the signal bandwidth of $B_s{\approx}N\Delta f$ and its $k^{\rm{th}}$ subcarrier is located at the frequency
%-------------------------------------------------------------------
\begin{equation}
{f_k} = {f_c} + k\Delta f,\,\,\,\,\,k =  - \frac{N}{2},...,\,\frac{N}{2} - 1,
\end{equation}
%-------------------------------------------------------------------
where $f_c$ is the center frequency. Let the nonoverlapping sets of data, pilot, and null subcarriers be defined as $S_D$, $S_P$, and $S_N$, respectively. Therefore, the transmitted passband analog signal in the time domain can be expressed as
%-------------------------------------------------------------------
%%-------------------------------------------------------------------
%\begin{equation}
%s(t) = 2{\text{Re}}\left\{ {\sum\limits_{k = 0}^{N - 1} {{S_k}{{\text{e}}^{j2\pi {f_k}t}}} p(t)} \right\},0 < t < {T_{{\text{bl}}}}
%\end{equation}
%%-------------------------------------------------------------------
%-------------------------------------------------------------------
\begin{equation}
s(t) = 2{\mathop{\rm Re}\nolimits} \left\{ {\sum\limits_{k \in {S_A}} {{s_k}{\kern 1pt} {{\rm{e}}^{j2\pi {f_k}t}}p(t)} } \right\},{\kern 1pt} {\kern 1pt} {\kern 1pt} {\kern 1pt} {\kern 1pt} 0 < t < {T_{\rm {bl}}}
\end{equation}
%-------------------------------------------------------------------
where $S_A=S_D\cup S_P$ represents the set of active subcarriers, $s_k$ is the modulated symbol on the $k^{\rm{th}}$ subcarrier, and $p(t)$ denotes the pulse shaping filter. Here, a rectangular window of length $T$ is used for pulse shaping.
%We direct the attention of the reader to \cite{Khodam_ICC} for more details on UWA channel.
%where $S_k$ is the modulated symbol on the $k^{\rm{th}}$ subcarrier and $p(t)$ is a pulse shaping filter.

\begin{figure}[t]
\centering
\includegraphics[width=.44\textwidth,height=48mm]{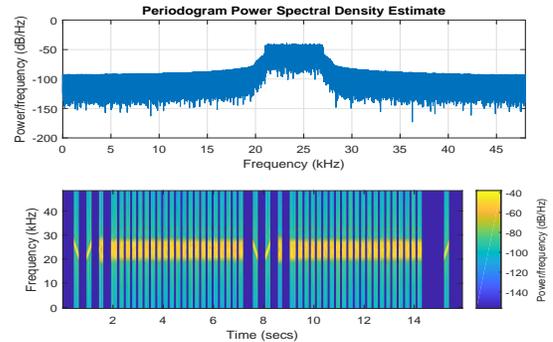}
\caption{PSD of the transmitted waveform in frequency band $[21-27]$ kHz.}
\label{fig:PSD_Waveform}
\vspace{-.25cm}
\end{figure}

The power spectral density (PSD) of the transmitted waveform is shown in~Fig.~\ref{fig:PSD_Waveform}. As depicted in~Fig.~\ref{fig:PSD_Waveform}, the preambles include a linear frequency-modulated (LFM) waveform, a hyperbolic frequency-modulated (HFM) waveform, an m-sequence coded waveform, and a cyclic-prefixed (CP) OFDM block~\cite{Zhou_UWA_Book_2014} to enable cross-correlation based signal detection. As it can be seen in Fig.~\ref{fig:PSD_Waveform}, following the preambles, there are twenty QPSK modulated OFDM blocks followed by another twenty OFDM blocks that is 16-QAM modulated. An HFM post-amble is appended to the end of the waveform, resulting in a 14.9-second total time duration of the waveform.
%-------------------------------------------------------------------
%-------------------------------------------------------------------
The synchronization and Doppler scale estimation are achieved through self-correlation of the CP-OFDM preamble. After synchronization, OFDM blocks are truncated and the symbols on the active subcarriers are obtained after the DFT module. A least squared (LS) estimator follows to estimate the channel with the help of pilot symbols. Here, a linear minimum mean squared error (LMMSE) equalizer is used for symbol detection. The detected symbols are then de-interleaved, and symbol-level soft metric is computed for the LDPC decoding module.

 %A least squared (LS) estimator follows to estimate the channel with the help of pilot symbols. Here, a linear minimum mean squared error (LMMSE) equalizer is followed by symbol detection and decoding module according to the modulation and coding schemes used. The parameters of the considered coding scheme are highlighted in the simulation results section.
% In case of the lake being frozen, the under-ice environment and the platform is quite stable. Therefore, there maybe no significant Doppler effects in the recorded waveforms and the corresponding Doppler compensation module could be taken off during the decoding process.

\section{MANP Design}\label{sec:Receiver Design}

The structure of the proposed receiver is shown in Fig.~\ref{fig:System Model}. The proposed MANP is implemented in the analog domain before the ADC. Since, locally optimum detection of signals in non-Gaussian noise exploits nonlinear kernels~\cite{Introduction_VincentPoor_1998}, the exact shape of the optimum kernel may be too complicated to be implemented by analog circuitry. Therefore, for easier implementation, a suboptimal threshold-based analog intermittent nonlinear preprocessor is proposed in this paper.

\begin{figure}[t]
\centering
\includegraphics[width=.36\textwidth,height=28mm]{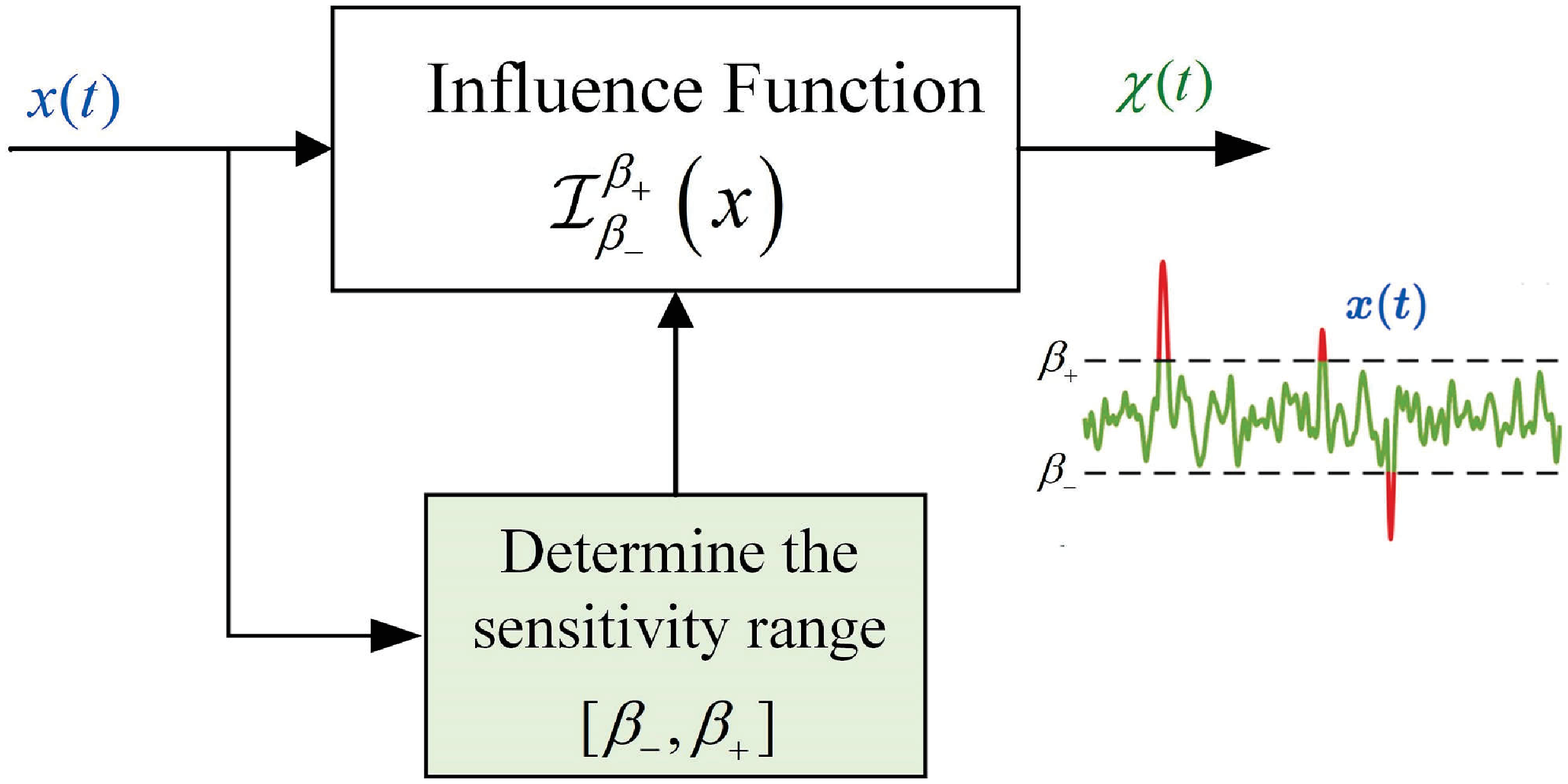}
\caption{Block diagram of generalized MANP.}
\label{fig:MANP_Diagram}
%\vspace{-.4cm}
\end{figure}

\begin{figure}[t]
\centering
\includegraphics[width=.41\textwidth,height=44mm]{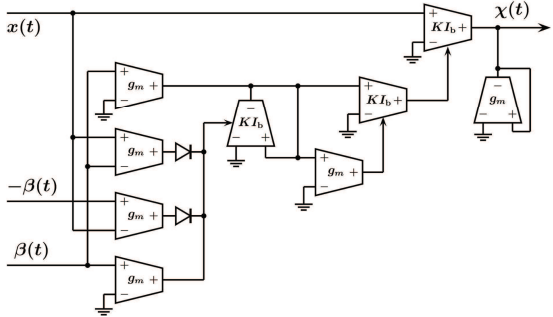}
\caption{Circuit for influence function with $\gamma=1$.}
\label{fig:MANP_Circuit}
\vspace{-.2cm}
\end{figure}

The general block diagram of MANP is shown in Fig.~\ref{fig:MANP_Diagram}. Here, $x(t)$ and $\chi(t)$ are the input and output of the MANP, respectively. The output of the MANP is represented as
%-------------------------------------------------------------------
\begin{equation}
\chi(t)=\InfFunc(x(t)),
\end{equation}
%-------------------------------------------------------------------
where $\InfFunc(x)$ is defined as the influence function. Note that, the behavior of MANP goes to the nonlinear regime in response to the amplitude of incoming outliers. Therefore, we will require that ${\InfFunc(x)}$ be effectively linear for ${\beta_-(t) \le x \le \beta_+(t)}$, and its absolute value monotonically decays to zero for $x$ outside of the range~$[\beta_-(t),\beta_+(t)]$. In general $|\beta_-(t)|$ and $|\beta_+(t)|$ are different, but for symmetric signals such as OFDM we can set $|\beta_-(t)|=|\beta_+(t)|=\beta(t)$. We refer to this $\beta(t)$ as the resolution parameter. Therefore, in practice we only need to find one resolution parameter $\beta(t)$ which determines the sensitivity range~$[-\beta(t),\beta(t)]$. For example, one realization of the influence function for MANP can be expressed as
%-------------------------------------------------------------------
\begin{equation}\label{eq:MANP}
\chi (t) = x(t)\left\{ \begin{array}{l}
1,\quad \quad \quad \quad \quad \,\, |x(t)| \le \beta (t)\\
{\left( {\frac{{\beta (t)}}{{\left| {x(t)} \right|}}} \right)^{\gamma  + 1}},\quad |x(t)| > \beta (t)
\end{array} \right.
\end{equation}
%-------------------------------------------------------------------
where $\gamma$ is a constant that determines how fast the proposed influence function transitions from clipping ($\gamma=0$) to blanking ($\gamma\to\infty$) and its value will differ based on the application (e.g., $\gamma=1$ is considered in this work). In other words, this influence function changes the nonlinearity from clipping to blanking based on the amplitude of incoming signal. Fig.~\ref{fig:MANP_Circuit} shows a practical schematic of MANP in \eqref{eq:MANP} with $\gamma=1$ based on the operational transconductance amplifiers (OTAs)~\cite{Parveen_OTA_2009}, which can be implemented in integrated circuit (IC). Here, $g_m$ denotes the transconductance; $I_b$ represents the current of the base in OTA unit; and $K$ is a constant with unit one over voltage.
%-------------------------------------------------------------------
%-------------------------------------------------------------------
%-------------------------------------------------------------------
%-------------------------------------------------------------------
\begin{figure}[t]
\centering
\includegraphics[width=.46\textwidth,height=46mm]{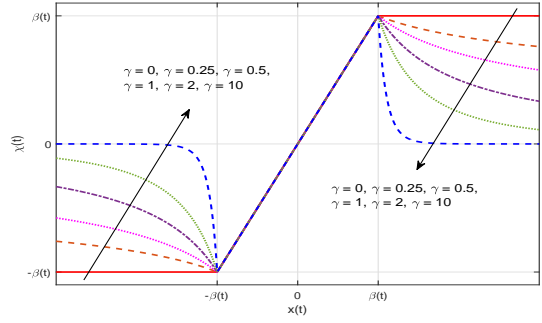}
\caption{Relation between input and output of MANP.}
\label{fig:MANP_InOut}
\vspace{-.3cm}
\end{figure}
%-------------------------------------------------------------------

The relationship between the input and the output of the MANP for different values of $\gamma$ is shown in Fig.~\ref{fig:MANP_InOut}. The expression in \eqref{eq:MANP} also demonstrates the disproportional behavior of the MANP on the signal of interest and impulsive noise. This nonlinear preprocessing increases the signal to noise ratio (SNR) in the desired frequency band by reducing the spectral density of the impulsive noise without significantly affecting the desired signal.

According to \eqref{eq:MANP}, the goal is to determine a proper resolution parameter $\beta(t)$ that enhances the quality of received signals under time-varying noise conditions. Therefore, an efficient value of $\beta(t)$ will maximize the suppression of the impulsive noise without distorting the signal of interest. Here, an effective value of the resolution parameter $\beta(t)$ is obtained as
%-------------------------------------------------------------------
\begin{equation}\label{eq:Tukey_Range}
\beta(t)=(1+2\beta_0)Q_2(t),
\end{equation}
%-------------------------------------------------------------------
where $Q_2(t)$ is the second quartile (median) of the absolute value of the input signal~$|x(t)|$, and $\beta_0$ is a constant coefficient (e.g. $\beta_0=1.5$). We direct the attention of the reader to \cite{Khodam_ICC} and \cite{Nikitin04adaptive_rank-filter} for more details on obtaining the quartile values in analog domain.

\section{Experimental Results}\label{sec:Experimental Results}

\begin{figure}[t]
\centering
\includegraphics[width=.38\textwidth,height=38mm]{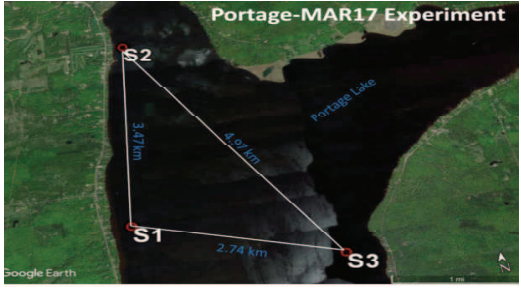}
\caption{Experiment setup.}
\label{fig:Testbed}
%\vspace{-.05cm}
\end{figure}

\begin{figure}
\centering
\includegraphics[width=.42\textwidth,height=35mm]{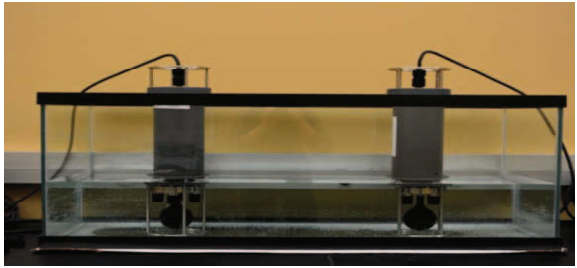}
\caption{OFDM Modem.}
\label{fig:OFDM_Modem}
\vspace{-.25cm}
\end{figure}

\begin{figure}[t]
\centering
\includegraphics[width=.48\textwidth,height=40mm]{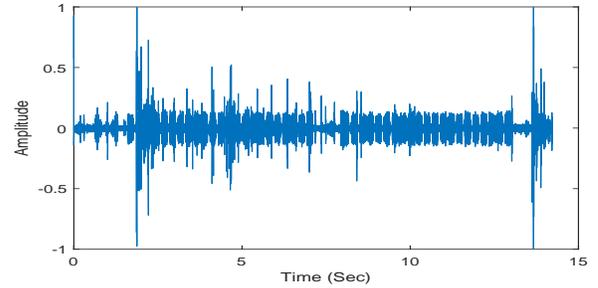}
\caption{Recorded OFDM waveform.}
\label{fig:WaveForm}
\vspace{-.25cm}
\end{figure}

On March 17, 2017, an under-ice experiment was conducted in Portage Lake, MI. The experimental setup is shown in~Fig.~\ref{fig:Testbed} and the OFDM modem that is used in this experiment is depicted in~Fig.~\ref{fig:OFDM_Modem}. During the experiment, the Portage Lake was covered by about $40$ cm thick ice. The water depth in the area varies from $8.3$ to $11.3$ meters. The transmitting node with an omnidirectional transducer was placed at $4.5$ meters below the water surface at $S_1$, as illustrated in~Fig.~\ref{fig:Testbed}. The receiving node with $4$-hydrophones was placed at $S_2$ at different depths and the transmission distance is $3.47$ km. An example of the recorded signal contaminated with impulsive noise at the receiver is depicted in~Fig.~\ref{fig:WaveForm}. For our numerical experiment the recorded signal was reconditioned for analog domain processing while retaining the measured characteristic of the impulsive noise. The system parameters of the considered OFDM system in UWA channels are listed in Table~\ref{tab:Simulation Parameters}. A total of 1024 subcarriers are used with 672 data subcarriers, 256 pilot subcarriers, and 96 null subcarriers. After the impulsive noise mitigation from the recorded signal, Doppler compensation and channel estimation can be done based on the measurements on null and pilot subcarriers, respectively. However, in this experiment the Doppler compensation module was taken off as the Doppler effect was negligible in the under ice situation. In the following, the SNR and BER performance are used to evaluate the performance of the proposed MANP in this experiment. In this paper we consider the time domain SNR which is obtained before the DFT module and can be expressed as
%-------------------------------------------------------------------
\begin{equation}\label{eq:SNR}
{\rm{{SNR}}} = \frac{{{P_s} - {P_n}}}{{{P_n}}}
\end{equation}
%-------------------------------------------------------------------
where $P_s$ and $P_n$ are the power of OFDM block and noise, respectively. The power of the OFDM block $P_s$ is considered as a summation of the desired signal power plus the noise power. The noise power $P_n$ can be obtained using the silence intervals in the waveform. For example, the intervals between preambles and the interval between the last OFDM block and the postamble. As long as the silence period is longer than the channel delay spread, there will be a clean portion without interference caused by the multipath effect.

\begin{figure}[t]
\centering
\includegraphics[width=.48\textwidth,height=48mm]{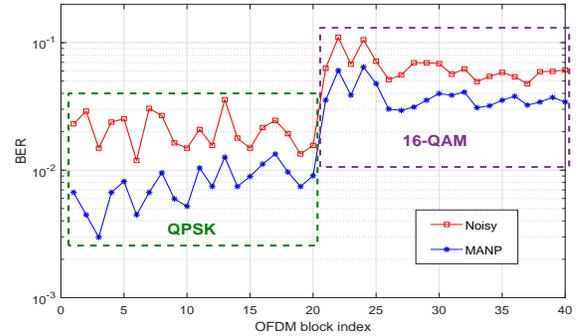}
\caption{BER of each OFDM block with and without MANP. }
\label{fig:BER_Block}
\vspace{-.25cm}
\end{figure}

\begin{figure}[t]
\centering
\includegraphics[width=.48\textwidth,height=48mm]{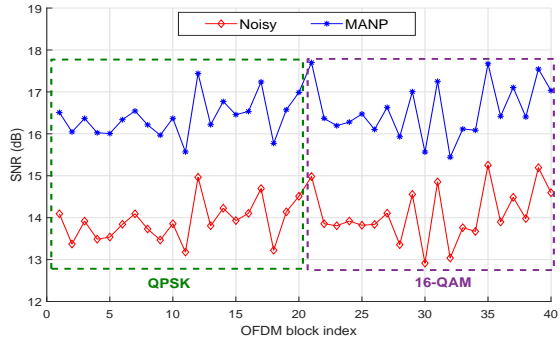}
\caption{Received SNR of each OFDM block with and without MANP.}
\label{fig:SNR_Block}
\end{figure}

The BER and SNR performance of each OFDM block in the receiver are shown in Fig.~\ref{fig:BER_Block} and Fig.~\ref{fig:SNR_Block}, respectively. Here, we just use the received signal from the first hydrophone but in general, the received signals by all the 4 hydrophones can be combined via the maximal ratio combining for joint decoding. As it can be seen in Fig.~\ref{fig:BER_Block} and Fig.~\ref{fig:SNR_Block}, the receiver performance is improved when the impulsive noise is suppressed by MANP. Without the outlier suppression, the power of the impulsive noise will spread over the entire frequency band of the OFDM block, which introduces error in the detection process.
%-------------------------------------------------------------------
\begin{table}[b]
\begin{center}
\captionsetup{labelfont=sc,labelsep=newline}
\caption{\sc{System Parameters}}
\begin{tabular}{ |l||c| }
\hline
\textbf{Parameters} & \textbf{Values} \\
\hline
\hline
 Modulation Scheme & QPSK-16-QAM\\
 Bandwidth ($B_s$) & 6 kHz \\
 Center Frequency($f_c$) & $24$ kHz \\
 No. of Subcarriers ($N$) & 1024 \\
 Subcarrier Spacing ($\Delta f$) & 5.88 Hz \\
 Sampling Frequency & $96$ kHz \\
 Symbol Duration ($T$) & 170.7 ms \\
 Guard Interval ($T_g$) & 79.3 ms \\
 Silence between preambles & 300 ms \\
 Silence between preamble and OFDM blocks& 100 ms \\
 LDPC Coding Rate (CR) & 1/2 \\
 Galois Field Size for QPSK and 16-QAM & GF(4), GF(16)\\
\hline
\end{tabular}
\label{tab:Simulation Parameters}
\end{center}
\end{table}
%-------------------------------------------------------------------
%-------------------------------------------------------------------
%-------------------------------------------------------------------
%-------------------------------------------------------------------
%-------------------------------------------------------------------
%-------------------------------------------------------------------
%-------------------------------------------------------------------
%-------------------------------------------------------------------
%--------------------------------------------------------------------
%----------------------------------------------------------------
%--------------------------------------------------------------------
\begin{figure}[t]
\centering
\includegraphics[width=.48\textwidth,height=48mm]{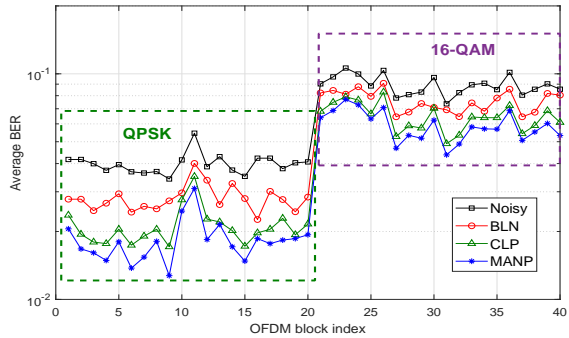}
\caption{Average BER of MANP, BLN, and CLP.}
\label{fig:BER_Avg_MANP_BLN_CLP}
\end{figure}
%----------------------------------------------------------------
%--------------------------------------------------------------------
\begin{figure}[t]
\centering
\includegraphics[width=.48\textwidth,height=48mm]{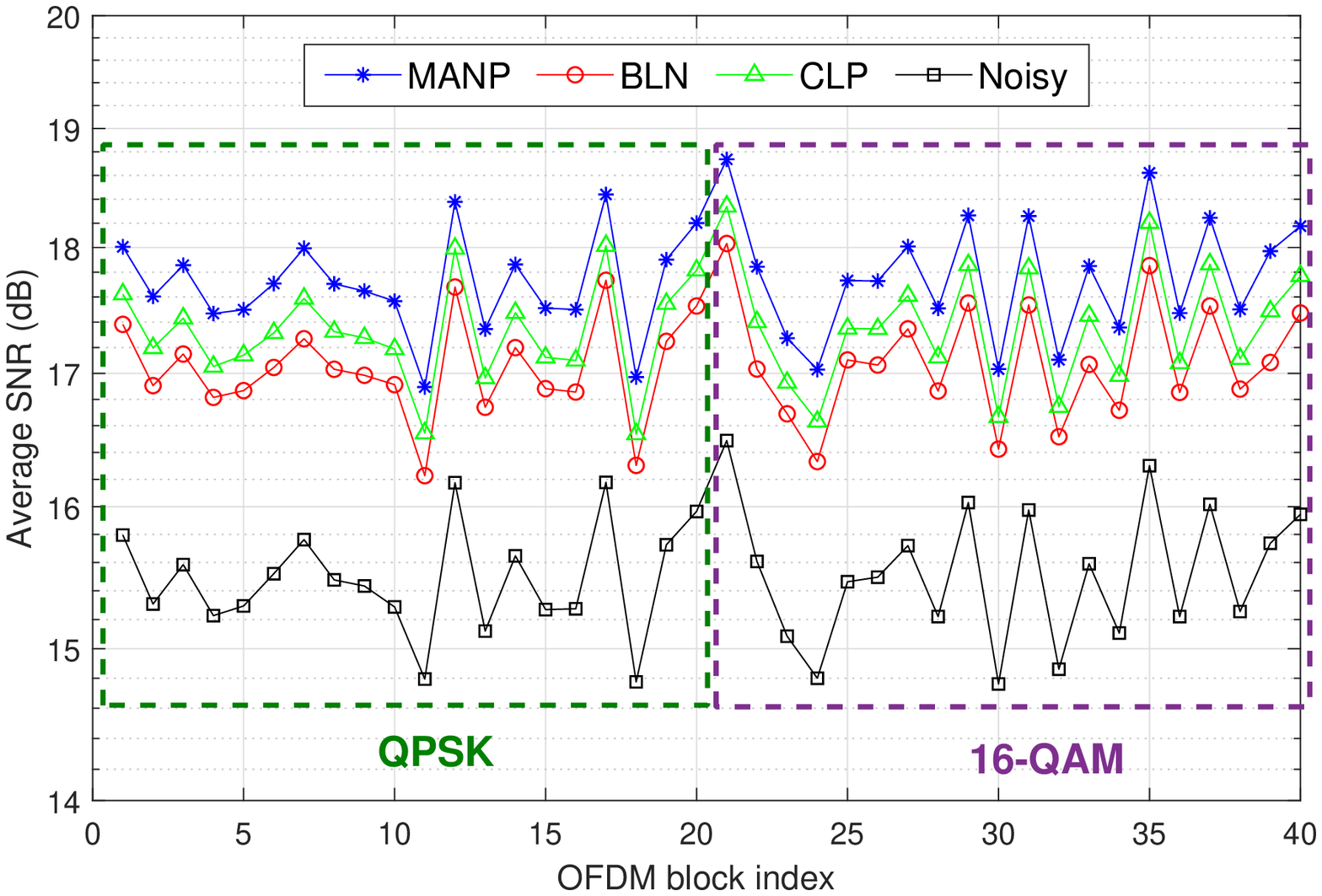}
\caption{Average received SNR of MANP, BLN, and CLP.}
\label{fig:SNR_Avg_MANP_BLN_CLP}
\end{figure}
%----------------------------------------------------------------
%--------------------------------------------------------------------
\begin{figure}[t]
\centering
\includegraphics[width=.48\textwidth,height=57mm]{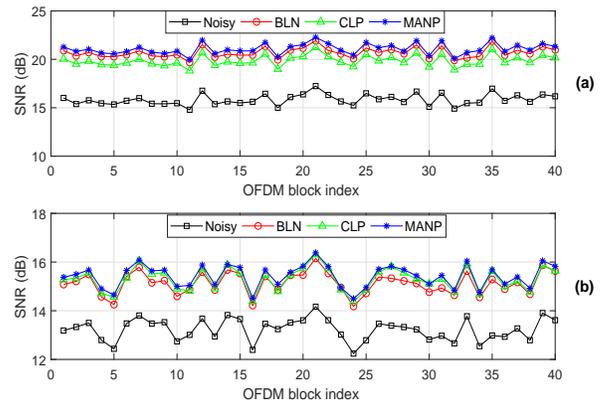}
\caption{Received SNR of MANP, BLN, and CLP.}
\label{fig:SNR_MANP_BLN_CLP}
\end{figure}
%----------------------------------------------------------------

In the following, we compare the performance of the MANP with two nonlinear digital approaches namely blanking (BLN) and clipping (CLP). Note that in all cases the thresholds for blanking and clipping are found according to the dynamic range of the received signal in the desired time window.

Fig.~\ref{fig:BER_Avg_MANP_BLN_CLP} and Fig.~\ref{fig:SNR_Avg_MANP_BLN_CLP} compare the average BER and the average received SNR of all three receivers, respectively. Here, the average is taken over ten recorded files. Fig.~\ref{fig:BER_Avg_MANP_BLN_CLP} shows that the BER performance of MANP outperforms both blanking and clipping in all investigated cases. The potency of MANP in reducing the power of impulsive noise in the signal passband is due to the fact that, unlike other nonlinear methods, MANP is implemented in the analog domain where the outliers are still broadband and distinguishable. As depicted in Fig.~\ref{fig:SNR_Avg_MANP_BLN_CLP}, the SNR with the proposed MANP surpasses both blanking and clipping in all studied cases. Fig.~\ref{fig:SNR_Avg_MANP_BLN_CLP} also shows that for our case studies, clipping outperforms blanking on average. However, in some cases blanking outperforms clipping~(Fig.~\ref{fig:SNR_MANP_BLN_CLP}-(a)) while in others clipping has better performance relative to blanking~(Fig.~\ref{fig:SNR_MANP_BLN_CLP}-(b)).

\section{Conclusions}\label{sec:Conclusion}

In this work, we proposed a novel memoryless analog nonlinear preprocessor (MANP) to alleviate the effect of impulsive noise in an OFDM-based UWA systems. The proposed MANP is implemented in the analog domain as the outliers are broadband and distinguishable. We also introduced a practical schematic of MANP based on the operational transconductance amplifiers (OTAs). Experimental results based on field data collected in an under-ice environment in Portage Lake, MI show that the proposed approach can provide significant improvement in the BER performance in the presence of strong impulsive components. In addition, the MANP-based approach outperforms other methods that use blanking or clipping for outlier suppression, especially at high levels of impulsivity.

\bibliographystyle{IEEEtran}

\bibliography{IEEEabrv,Reference}

\end{document}